\documentclass[sigconf]{acmart}
\usepackage{enumitem}
\usepackage{algorithm2e}
\RestyleAlgo{ruled}
\usepackage{multirow}
\usepackage{url}
\usepackage{listings, xcolor}

\newcommand{\code}[1]{\begin{small}\texttt{#1}\end{small}}

\definecolor{codegreen}{rgb}{0,0.6,0}
\definecolor{codepurple}{rgb}{0.58,0,0.82}
\definecolor{backcolour}{rgb}{0.95,0.95,0.92}

\lstdefinestyle{mystyle}{
    commentstyle=\color{codegreen},
    keywordstyle=\color{blue},
    stringstyle=\color{codepurple},
    basicstyle=\ttfamily\scriptsize,
    breakatwhitespace=false,
    breaklines=true,
    captionpos=b,
    keepspaces=true,
    showspaces=false,
    showstringspaces=false,
    showtabs=false,
    tabsize=2
}

\lstdefinestyle{mystyle-border}{
    commentstyle=\color{codegreen},
    keywordstyle=\color{blue},
    stringstyle=\color{codepurple},
    basicstyle=\ttfamily\scriptsize,
    breakatwhitespace=false,
    breaklines=true,
    captionpos=b,
    keepspaces=true,
    showspaces=false,
    showstringspaces=false,
    showtabs=false,
    tabsize=2,
    frame=single,
}

\lstdefinestyle{prompt}{
        basicstyle=\ttfamily\scriptsize,
        language=Html,
        commentstyle=\color{linenumbergray},
        stringstyle=\color{javapurple},
        keywordstyle=\color{red},
        morekeywords={@Test},
        morecomment=[s][\color{linenumbergray}]{/**}{*/},
        numberstyle=\tiny\color{linenumbergray},
        numbersep=2.5pt,
        xleftmargin=1em,
        moredelim=**[is][\color{javapurple}]{@h@}{@h@},
        morecomment=[f][{\btHL[fill=gitdel]}]-,
        morecomment=[f][{\btHL[fill=gitadd]}]+,
        breaklines = true,
        frame = single
}

\newcommand{\name}{CASCADE}

\AtBeginDocument{%
  }

\copyrightyear{2026}
\acmYear{2026}
\setcopyright{cc}
\setcctype{by}
\acmConference[ICSE-SEIP '26]{2026 IEEE/ACM 48th International Conference on Software Engineering}{April 12--18, 2026}{Rio de Janeiro, Brazil}
\acmBooktitle{2026 IEEE/ACM 48th International Conference on Software Engineering (ICSE-SEIP '26), April 12--18, 2026, Rio de Janeiro, Brazil}
\acmPrice{}
\acmDOI{10.1145/3786583.3786873}
\acmISBN{979-8-4007-2426-8/2026/04}

\begin{document}

\title{\name: LLM-powered JavaScript Deobfuscator at Google}

\author{Shan Jiang}
\authornote{Work done while at Google.}
\affiliation{%
  \institution{The University of Texas at Austin}
  \city{Austin}
  \country{USA}}
\email{shanjiang@utexas.edu}

\author{Pranoy Kovuri}
\authornotemark[1]
\affiliation{%
  \institution{Google}
  \city{Sunnyvale}
  \country{USA}}
\email{kovuripranoy@gmail.com}

\author{David Tao}
\affiliation{%
  \institution{Google}
  \city{Sunnyvale}
  \country{USA}
}
\email{dtao@google.com}

\author{Zhixun Tan}
\affiliation{%
 \institution{Google}
 \city{Sunnyvale}
  \country{USA}
}
\email{tzx@google.com}


\newcommand{\Comment}[1]{}

\begin{abstract}
Software obfuscation, particularly prevalent in JavaScript, hinders code comprehension and analysis, posing significant challenges to software testing, static analysis, and malware detection. This paper introduces \name, a novel hybrid approach that integrates the advanced coding capabilities of Gemini with the deterministic transformation capabilities of a compiler Intermediate Representation (IR), specifically JavaScript IR (JSIR). By employing Gemini to identify critical prelude functions—the foundational components underlying the most prevalent obfuscation techniques—and leveraging JSIR for subsequent code transformations, \name~effectively recovers semantic elements like original strings and API names, and reveals original program behaviors. This method overcomes limitations of existing static and dynamic deobfuscation techniques, eliminating hundreds to thousands of hard-coded rules while achieving reliability and flexibility. \name~is already deployed in Google’s production environment, demonstrating substantial improvements in JavaScript deobfuscation efficiency and reducing reverse engineering efforts.
\end{abstract}



\maketitle

\section{Introduction}

\begin{figure*}[htbp]
\begin{lstlisting}[language=java, style=mystyle-border]
function hi() {
  console.log('Hello World!');
}
hi();
\end{lstlisting}
\begin{lstlisting}[language=java, style=mystyle-border]
function _0x432d() {
   var _0x1398fd = [
       '754705BaCmnb', '710BeTfxi', '6078JFPbmg', '391457WTcWXy', '6916iSWoZO',
       '1533357iISYWF', '2400oHfZQf', '24eeZIfI', 'log', '149YcoFBV',
       '1367628nfkDqA', '1243948epsCBk', 'Hello\x20World!'
   ];
   _0x432d = function () {
       return _0x1398fd;
   };
   return _0x432d();
}
(function (_0x38057e, _0xee6281) {
   var _0x18e9b0 = _0x4c0c;
   var _0x1ab84e = _0x38057e();
   while (!![]) {
       try {
           var _0xc7d52d = parseInt(_0x18e9b0(0x1b7)) / 0x1 * (-parseInt(_0x18e9b0(0x1bf)) / 0x2) + -parseInt(_0x18e9b0(0x1b8)) / 0x3 + -parseInt(_0x18e9b0(0x1b9)) / 0x4 + -parseInt(_0x18e9b0(0x1c1)) / 0x5 * (parseInt(_0x18e9b0(0x1bd)) / 0x6) + parseInt(_0x18e9b0(0x1bb)) / 0x7 * (-parseInt(_0x18e9b0(0x1b5)) / 0x8) + -parseInt(_0x18e9b0(0x1c0)) / 0x9 + -parseInt(_0x18e9b0(0x1bc)) / 0xa * (-parseInt(_0x18e9b0(0x1be)) / 0xb);
           if (_0xc7d52d === _0xee6281) {
               break;
           } else {
               _0x1ab84e['push'](_0x1ab84e['shift']());
           }
       } catch (_0x5a174b) {
           _0x1ab84e['push'](_0x1ab84e['shift']());
       }
   }
}(_0x432d, 0x40942));
function _0x4c0c(_0x32b956, _0x514a26) {
   var _0x432d26 = _0x432d();
   _0x4c0c = function (_0x4c0c62, _0x284b04) {
       _0x4c0c62 = _0x4c0c62 - 0x1b5;
       var _0x85349e = _0x432d26[_0x4c0c62];
       return _0x85349e;
   };
   return _0x4c0c(_0x32b956, _0x514a26);
}
function hi() {
   var _0x964834 = _0x4c0c;
   console[_0x964834(0x1b6)](_0x964834(0x1ba));
}
hi();
\end{lstlisting}
\caption{Hello World obfuscated by Obfuscator.IO with the default (the lowest level) configuration}
\Description{An example of obfuscated code generated by Obfuscator.IO with the default which is the lowest level of configuration}
\label{fig:default_obfuscation}
\end{figure*}

JavaScript is the dominant programming language for web development, powering client-side interactions across billions of web pages, mobile applications, and browser extensions. However, its widespread adoption has led to increased use of code obfuscation techniques that deliberately transform readable code into complex, difficult-to-understand variants. This poses significant challenges to software testing, complicates analysis, and hinders malware detection. Deobfuscation, which transforms the code to restore readability and reveal the original intention while maintaining code semantic equivalence, is inherently difficult. It requires robust handling of various complex obfuscation techniques—such as dynamic code generation (e.g., using the eval function to execute code from strings), control flow flattening, and string encoding. Deobfuscation involves not only syntactic simplification but also deeper semantic restructuring to reverse transformations like identifier mangling, opaque predicate removal, and constant unfolding. Aggressive code transformations might significantly enhance readability but inadvertently compromise semantic equivalence; conservative approaches may produce code that, while correct, remains largely inscrutable.

Based on empirical observations, Obfuscator.IO \cite{obio} is the most widely used JavaScript obfuscator by malware developers, with js-confuser \cite{js-confuser} is the second most common. Obfuscator.IO's level of obscurity, as demonstrated in a concrete example shown in Fig \ref{fig:default_obfuscation}, significantly impedes malware detection workflows. Amongst its various obfuscation techniques, the obfuscation of strings and API names (method and property names) creates the greatest barriers to code analysis. Recovering original string literals and API names would markedly reduce the manual effort required in Google’s malware detection processes. Since Obfuscator.IO performs API obfuscation by transforming direct API calls (e.g., \code{chrome.cookies}) as string-indexed lookups (\code{chrome["cookies"]}) before obfuscating the string literal (\code{"cookies"}), solving string obfuscation inherently resolves the API obfuscation. Consequently, we target Obfuscator.IO’s string obfuscation in this paper.

To address string obfuscation of Obfuscator.IO, we propose \name~(\textbf{C}ombined \textbf{A}nalysis of \textbf{S}cripts with a \textbf{C}ontext-\textbf{A}ware \newline \textbf{D}eobfuscation \textbf{E}ngine), a hybrid JavaScript deobfuscator that combines large language models (LLMs) and a compiler intermediate representation (IR). First, it uses Gemini \cite{comanici2025gemini} to detect and extract key code patterns, termed \textit{prelude functions}, generated by Obfuscator.IO for manipulating obfuscated strings. Then, it runs a customized constant propagation and inline pass built on JSIR \cite{jsir}, a next-generation compiler framework, and treats prelude functions as pure (i.e. idempotent and side-effect-free) functions and dynamically execute them in a sandboxed JavaScript environment. The use of LLMs for code pattern detection eliminates reliance on manually engineered heuristics; the compiler IR provides a robust structure for applying code transformations, ensuring functional integrity through semantics-preserving operations while systematically improving code readability.

Recent work confirms that LLMs possess strong code understanding skills \cite{li2023cctest,nam2024using,hong2025effectiveness}, making them promising tools for identifying code patterns. LLMs also align user prompts with their learned internal representations \cite{jiang2024generating} and show emergent competence on tasks not seen during training \cite{schaeffer2024emergent}. Together, these properties position LLMs to automatically pinpoint prelude functions in long, obfuscated codebases. However, LLMs still falter on tasks that demand exact logical or mathematical reasoning \cite{zhao2024docmath,yuan2023well} limiting their effectiveness in fully correct end-to-end deobfuscation. Even small arithmetic mistakes—e.g., flipping an if condition from true to false—can radically change program behavior. Evaluating pure LLM deobfuscation is also difficult because proving functional equivalence between the obfuscated and recovered programs is non-trivial. The hybrid approach of \name~leverages the best of both worlds: Gemini detects prelude functions with a 99.56\% average success rate on a synthetic dataset of 12K obfuscated files; the deterministic transformations of JSIR then restores 945 string literals per sample in an average of two seconds, demonstrating practical throughput.

This paper offers three primary contributions:

\begin{itemize}
    \item Novelty — \name~is the first framework to pair an LLM with compiler-level IR transformations, merging probabilistic code understanding with deterministic rewrites for robust deobfuscation of JavaScript code.
    \item Industrial deployment — \name~is deployed in production at Google and eliminates the need for hard-coded pattern matching rules (100+ lines per obfuscation technique) in WebCrack and Deobfuscator.IO.
    \item Robustness - \name~hybrid architecture eliminates hard-coded pattern matching which can be easily broken with just a slight change after obfuscation, and is robust to regression changes of Obfuscator.IO.
    \item Scalability — We open-source our prompt templates and the full JSIR infrastructure to facilitate community adoption and reproducibility. The repository is publicly available at \url{https://github.com/google/jsir}.
\end{itemize}

\section{Background and Related Work}
\subsection{Software Obfuscation.} 
Software obfuscation deliberately transforms code to hinder readability, analysis, and reverse engineering. Common obfuscation techniques include restructuring code, replacing descriptive variable names with unintuitive identifiers, injecting redundant or misleading instructions, manipulating control flow, and encrypting literals such as strings or configuration data \cite{6185286,zhang2021android}. Although obfuscation can legitimately safeguard proprietary algorithms and sensitive resources (e.g., IP addresses) \cite{doyle2018privacy,lynn2004positive}, it is also exploited by attackers to mask malicious logic—especially in web and mobile scripts \cite{brezinski2023metamorphic,281380,10.1016/j.cose.2015.02.007}. 

Obfuscated code severely diminishes the effectiveness of analysis tools, creating a major obstacle for software testing and static analysis methods \cite{10.1145/3597926.3598061,10.1109/ICSE-C.2017.79,10.1145/3293882.3330563}. Moreover, obfuscation hampers malware detection by concealing malicious behavior from static analyzers, security filters, machine-learning detectors, and manual reviewers \cite{pantelaios2024fv8,ren2023jsrevealer,Li2018JSgraphER}. JavaScript, the dominant client-side language, is particularly vulnerable because its source code is delivered directly to the browser \cite{fraunholz2018defending,10.5555/2028067.2028070}, making it an attractive target for both defensive and malicious obfuscation.

\subsection{JavaScript Deobfuscation}
The prevalence of malicious JavaScript code underscores an urgent need for effective deobfuscation techniques \cite{10.1145/3597926.3598146,10.1145/3691620.3695492,chen2025jsdeobsbench}. JavaScript deobfuscators aim to restore code readability for analysis while preserving semantic correctness. Various deobfuscation approaches have been explored in recent years.

Machine learning-based approaches, exemplified by DEGUARD \cite{statisticalDeob} for ProGuard-obfuscated Android code, have demonstrated promising results. However, the inherent stochasticity of current ML models prevents them from guaranteeing semantic equivalence between original and deobfuscated code.

Dynamic analysis can enhance code readability to facilitate manual inspection of obfuscated scripts \cite{herrera2020optimizing}. Lu et al. \cite{lu2012automatic} proposed a dynamic analysis technique incorporating program slicing. However, dynamic methods generally impose specific runtime environment requirements, incur substantial performance overhead, and raise security concerns.

Static analysis presents an alternative approach. JSDES \cite{abdelkhalek2017jsdes} introduced function-centric deobfuscation but struggles with obfuscations implemented purely through basic operations. This limitation renders it less effective against Obfuscator.IO, which commonly interleaves such operations with complex function calls. Many static techniques operate at the AST level \cite{10.1145/3319535.3345656,releim,jsdeob,webcrack}. TransAST \cite{qin2023transast} uses static analysis and machine translation to deobfuscate JavaScript but faces difficulties with obfuscation techniques that are dynamically generated or manifested at runtime.

Pattern-matching approaches, such as Safe-Deobs \cite{herrera2020optimizing}, rely on predefined patterns derived from real-world malware. Webcrack \cite{webcrack} is a rule-based JavaScript deobfuscator that transforms code at the AST level. Webcrack relies on hard-coded AST rules, so even slight modifications (e.g., changing \code{while (!![])} to \code{while (!false))} can prevent successful deobfuscation. AST-based methods are limited when encountering heavily obfuscated code or novel obfuscation patterns not included in their predefined rule sets. The key limitation of AST-based tools is their lack of semantic understanding capabilities compared to compiler-based methods. These tools are prone to altering code behavior or introducing errors during the deobfuscation process.

\subsection{LLM for SE.}
Large Language Models (LLMs) have recently demonstrated strong capabilities in diverse code-related tasks, including program synthesis, code refactoring, and automated test generation \cite{jiang2024generating,zhong2025approach,li2023cctest,zhong2025april}. Trained on extensive corpora of code and textual data, these models develop an implicit understanding of programming language syntax, semantics, and code structures \cite{liu2024llm,richards2024you,jiang2025obsmith}. Consequently, LLMs exhibit notable capabilities in tasks requiring both natural language comprehension and code manipulation \cite{jiang2024generating}. While models such as StarCoder \cite{lozhkov2024starcoder} and Gemini demonstrate proficiency in code generation, they remain susceptible to hallucination, wherein they generate plausible but incorrect or nonsensical outputs \cite{gambardella-etal-2024-language}. Moreover, LLMs exhibit limitations in tasks demanding precise logical and mathematical reasoning \cite{zhao2024docmath,yuan2023well}. This constrains their efficacy in achieving correct end-to-end deobfuscation, where precise computation is crucial, as even minor inaccuracies can drastically alter program behavior. To address these non-deterministic limitations, hybrid approaches integrating LLMs with Abstract Syntax Trees (ASTs) or Intermediate Representations (IRs) have demonstrated promise in various software engineering applications \cite{nikolov2025google}.

\section{String Obfuscation And \name~Approach}
\begin{figure}[htbp]
    \centering
    \includegraphics[width=\linewidth]{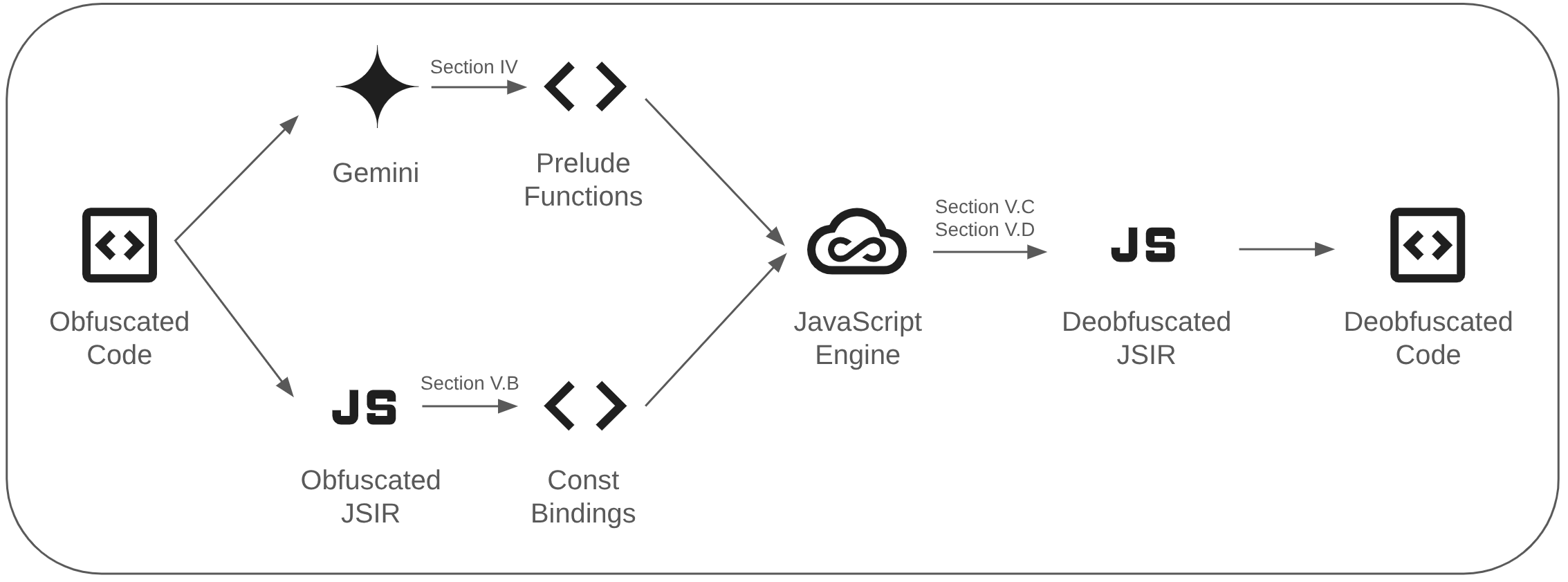}
    \caption{\name~Deobfuscator}
    \Description{Overall architecture of CASCADE, we combine LLMs and traditional compiler IR (JSIR) in our workflow}
    \label{approach}
\end{figure}
Obfuscator.IO represents a widely adopted obfuscation tool among malicious actors \cite{nowyousee}. The tool employs a multi-stage obfuscation pipeline incorporating string obfuscation, control flow flattening, and complex arithmetic operations, etc.. String obfuscation emerges as the most critical technique for two key reasons: (1) recovering string literals and API names yields the greatest readability improvements, and (2) string obfuscation occurs as the final pipeline stage, layering atop other obfuscation methods and establishing string deobfuscation as a prerequisite for subsequent analysis.
\textbf{Consequently, \name~prioritizes string deobfuscation when processing Obfuscator.IO-generated code.}

The Obfuscator.IO tool implements string obfuscation through two code transformations. First, it creates a \textbf{string fetching function} that retrieves and optionally decodes elements from a \textbf{global string table}. Second, it replaces each original string literal with a complex \textbf{string recovery expression} that calls the fetching function through multiple abstraction layers including wrapper functions, object property lookups, variable aliases, and arithmetic operations.

Figure \ref{fig:sim_string_obfuscation} is a simplified version of Figure \ref{fig:default_obfuscation}, and illustrates string obfuscation in Obfuscator.IO. In this example, the strings \code{"log"} and \code{"Hello World!"} are replaced by calls to a \code{getString()} function, which retrieves the corresponding entries from a global string array using offset indices.

\begin{figure}[htbp]
\begin{lstlisting}[language=java, style=mystyle]
function hi() {
  console.log('Hello World!');
}
hi();
\end{lstlisting}
\hrule
\begin{lstlisting}[language=java, style=mystyle]
var globalStringTable = [
  ..., /*index: 8*/'log',
  ..., /*index: 12*/'Hello World!'
];
function getStringArray() {
  return globalStringTable;
}

function getString(index) {
  return getStringArray()[index - 437];
}

(function () {
  var stringArray = getStringArray();
  while (true) {
    if (conditionOnPermutationOf(stringArray)) {
      stringArray.push(stringArray.shift());
    }
  }
})();

function hi() {
  console[getString(438)](getString(442));
}
hi();
\end{lstlisting}
\caption{Simplified Illustration of String Obfuscation}
\Description{A real example of string obfuscation in Obfuscator.IO}
\label{fig:sim_string_obfuscation}
\end{figure}

The global string table undergoes rotation an unknown number of times, which makes \code{getStringArray()}—and by extension, \code{getString()}—seem non-idempotent with side effects, despite their truly idempotent and side-effect-free nature. As a result, these functions bypass compiler optimizations such as constant propagation and inlining.

However, the definition and initialization of the global string table, along with the string fetching function—which we term \textbf{prelude functions}—are generated from templates and are highly recognizable. By detecting these code segments, treating \code{getString()} as an idempotent and side-effect-free built-in function, and dynamically executing it upon invocation, we can enhance constant propagation and inlining, thereby successfully evaluating the string recovery expressions.

To achieve this, the \name~deobfuscator employs a hybrid approach that integrates static analysis, dynamic execution, and AI:

\begin{enumerate}[label=(\arabic*)]
  \item We identify prelude functions by prompting Gemini, which leverages its high-level code understanding and structured output capabilities.

  \item We dynamically execute these detected prelude functions in a sandboxed JavaScript environment to obtain accurate results from string fetching function calls.

  \item We apply constant propagation and inlining via JSIR—a JavaScript intermediate representation built on MLIR \cite{mlir}—to evaluate arithmetic expressions and resolve indirections.
\end{enumerate}

This hybrid approach has the following key benefits:

\begin{enumerate}[label=(\arabic*)]

\item \textbf{Hybrid dynamic execution}: \name~overcomes the conservatism of pure static analysis, which often fails against complex obfuscations to maintain soundness. By dynamically executing code snippets that appear non-idempotent but are actually idempotent, \name~evaluates more expressions and recovers obfuscated strings effectively. 

\item \textbf{AI-driven maintainability}: \name~eliminates the need for hundreds to thousands of lines of brittle, unreadable manual rules in prelude function detection by leveraging state-of-the-art Gemini, achieving accuracy of 99.56\%. Such manual rules are easy to break with minor code changes (e.g. change \code{true} to \code{!false}, or make an alias to a variable), as shown in Figure \ref{fig:minor_changes}. In comparison, Gemini is resilient against minor changes in code format and syntax, which evades traditional detection based on AST or regex rules.

\item \textbf{Advanced static analysis via JSIR}: \name~conducts comprehensive code analyses and transformations using the robust JSIR infrastructure, a novel high-level JavaScript intermediate representation (IR). Although ASTs preserve high fidelity to source code syntax and suit source-to-source transformations \cite{10.1145/2950290.2950308,ndichu2019machine}, IRs facilitate more sophisticated semantic-level analyses and transformations \cite{szafranieccode,jiset}. JSIR uniquely encodes all AST syntactical information while supporting dataflow analysis.

\item \textbf{Responsible use of AI}: \name~deliberately refrains from using LLMs to directly generate deobfuscated code, eliminating a wide range of potential hallucination errors. Its hybrid design improves explanability, observability, and evaluability, ensuring practical use in production environments.

\end{enumerate}

\begin{figure*}
	\centering

	\begin{minipage}[t]{0.37\linewidth}
\begin{lstlisting}[language=java, style=mystyle]
function _0x4d08() {
  var _0x235313 = [ ... ];

  _0x4d08 = function () {
    return _0x235313;
  };
  return _0x4d08();
}

while (true) { ... }

while (!![]) { ... }

while (true) { ... }
\end{lstlisting}
	\end{minipage}
	\begin{minipage}[t]{0.44\linewidth}
		\begin{lstlisting}[language=java, style=mystyle]
function _0x4d08() {
  var _0x235313 = [ ... ];
  var _alias = _0x235313; // Add an alias
  _0x4d08 = function () {
    return _alias;
  };
  return _0x4d08();
}

while (!false) { ... } // true => !false

while (!!true) { ... } // !![] => !!true

for (; !false; ) { ... } // Another infinite loop
\end{lstlisting}
	\end{minipage}

 \caption{Slight changes cause rule-based pattern match to fail, left means Webcrack deobfuscation is \textbf{correct}, right means Webcrack \textbf{cannot} deobfuscate}
 \Description{Some slight changes which cause rule-based pattern match to fail. Left part means Webcrack deobfuscation is \textbf{correct}, right part means Webcrack \textbf{cannot} deobfuscate}
		\label{fig:minor_changes}
 
\end{figure*}

\section{Detecting prelude functions}
\label{sec:prelude}

\begin{figure*}[htbp]
\begin{lstlisting}[language=java, style=prompt]
Obfuscator.io is a well-known JavaScript obfuscator.
To obfuscate strings, it usually inserts 3 templates: StringArrayTemplate, StringArrayRotateFunctionTemplate and
StringArrayCallsWrapperTemplate.
## StringArrayTemplate
StringArrayTemplate defines a string array that includes string literals from the original unobfuscated code.
It's generated using this template:
```js
{{string_array_template}}
```
An example looks like this:
```js
{{string_array_template_example}}
```
## StringArrayCallsWrapperTemplate
StringArrayCallsWrapperTemplate fetches a string from the string array given a shifted index.
Depending on the obfuscation config, StringArrayCallsWrapperTemplate may also perform custom decoding such as base64 or rc4.
It's generated using this template:
```js
{{string_array_calls_wrapper_template}}
```
### StringArrayCallsWrapperTemplate example 1:
```js
{{string_array_calls_wrapper_template_example_1}}
```
### StringArrayCallsWrapperTemplate example 2:
...
### StringArrayCallsWrapperTemplate example 3:
...
## StringArrayRotateFunctionTemplate
StringArrayRotateFunctionTemplate is a piece of code that rotates the string array defined in StringArrayTemplate.
It's generated using this template:
```js
{{string_array_rotate_function_template}}
```
An example looks like this:
```js
{{string_array_rotate_function_template_example_1}}
```
Another example looks like this:
```js
{{string_array_rotate_function_template_example_2}}
```
Now, as an obfuscator.io expert, your job is to detect these templates from a piece of obfuscated code.
In particular, the code is split into N parts, using html tags in comments with an ID.
Please return a JSON map from template name to a list of parts belonging to that template.
Here is an example. If the input is this:

<example_java_script>
// <0>
var _0x42f9c1 = _0x1e0f;
// </0>
// <1>
String Array Rotate Function ...
// </1>
// <2>
console[_0x42f9c1(337)]('Hello, world!');
// </2>
// <3>
String Fetching Function ...
// </3>
// <4>
String Array Function ...
// </4>
</example_java_script>

Output:
{
  'StringArrayTemplate': 4,
  'StringArrayRotateFunctionTemplate': 1,
  'StringArrayCallsWrapperTemplate': 3
}

Now please investigate the following code and return the output JSON.
<investigation_java_script>
{{annotated_example}}
</investigation_java_script>
**Remember:**
* The values for StringArrayTemplate, StringArrayRotateFunctionTemplate, and StringArrayCallsWrapperTemplate can never be equal because of the nature of obfuscation.
* Only output the json, do not output any explanation

Output: 

\end{lstlisting}
\caption{Prompt Template For Prelude Detection}
\Description{Prompt used for prelude funtion detection.}
\label{fig:prompt}
\end{figure*}

Prelude functions, generated from templates, are easily recognizable to reverse engineers despite obfuscation efforts. However, manual detection becomes prohibitively time-consuming when applied to lengthy obfuscated codebases. Rule-based approaches using AST patterns \cite{webcrack} or regular expressions \cite{obio} prove equally problematic—they require extensive manual crafting, demand ongoing maintenance, and exhibit brittleness when confronted with minor code variations.

Recent advances in LLM code comprehension capabilities, exemplified by models such as Gemini, motivated our investigation into LLM-based prelude detection. The remainder of this section is organized as follows: Section \ref{sec:prelude_description} presents a comprehensive analysis of prelude function characteristics, Section \ref{sec:prompt} details our prompting methodologies, and Section \ref{sec:cost} outlines optimization strategies for cost reduction and error mitigation.

\subsection{Prelude functions}
\label{sec:prelude_description}
Obfuscator.IO implements string obfuscation through three prelude functions:

\begin{enumerate}[label=(\arabic*)]

\item A string array function that defines a global string table as an array.

\item A string fetching function that retrieves elements from the array using shifted indices.

\item A string array rotate function implemented as an IIFE (immediately invoked function expression) that rotates the global string table until a complex arithmetic expression evaluates to a target value.

\end{enumerate}

The following examples illustrate these prelude functions using simplified names for clarity. In practice, Obfuscator.IO generates incomprehensible function names such as \code{\_0x746cd9}.

\textbf{String Array Function}: Figure \ref{fig:string_array_func} demonstrates the \\ \code{getStringArray()} function, which returns a reference to a global string table containing both original string literals and those generated during obfuscation. This function employs a self-redefining pattern: upon first invocation, it redefines itself to return a reference to the string array created during that initial call. JavaScript closure mechanisms ensure that \code{getStringArray()} consistently returns the same array object in subsequent invocations.

\begin{figure}[htbp]
\begin{lstlisting}[language=java, style=mystyle]
function getStringArray() {
  var stringArray = [
      'info', 'ewLHr', 'prototype', ...,
      'ctor(\x22retu', 'Hello\x20Worl',
      'tZhir', '__proto__', 'table'
  ];
  getStringArray = function () {
    return stringArray;
  };
  return getStringArray();
}

\end{lstlisting}
\caption{String Array Function (Variable Renamed For Readability)}
\Description{An example of string array function.}
\label{fig:string_array_func}
\end{figure}

\textbf{String Fetching Function}: Figure \ref{fig:string_fetching_func} demonstrates the \\ \code{getString()} function, which retrieves a single element from the global string table using an index-based lookup with a fixed offset. Similar as \code{getStringArray()}, the \code{getString()} function employs self-modification during its initial execution, rendering it non-idempotent. This behavior likely serves as an anti-analysis technique designed to prevent compiler optimization through inlining. Figure 3 illustrates the default obfuscation level implementation. At elevated obfuscation levels, the global string table contains encoded strings, and \code{getString()} runs custom decoding algorithms (such as Base64 or RC4 decryption), significantly increasing the function's complexity and length.

\begin{figure}[htbp]
\begin{lstlisting}[language=java, style=mystyle]
function getString(index, ignored) {
  var stringArray = getStringArray();
  return getString = function (index, ignored) {
    index = index -
            (-0x3*0x23b + 0x1b3 + -0x1*-0x676);
    var result = stringArray[index];
    return result;
  }, getString(index, ignored);
}

\end{lstlisting}
\caption{String Fetching Function (Variable Renamed For Readability)}
\Description{An example of string fetching function.}
\label{fig:string_fetching_func}
\end{figure}

\textbf{String Array Rotate Function}: Figure \ref{fig:string_rotate_func} demonstrates an IIFE that systematically rotates the global string table until a specific mathematical expression evaluates to a target value. The expression applies parseInt() to selected table strings and performs arithmetic operations, making the evaluation dependent on the table's current permutation order. Achieving the correct permutation is essential for getString() to successfully retrieve the intended string value.

\begin{figure}[htbp]
\begin{lstlisting}[language=java, style=mystyle]
(function (getStringArray, target) {
  var stringArray = getStringArray();
  function getStringWrapper1(a, b, c, d) {
    return getString(d - 0x1e2, a);
  }
  ...
  while (!![]) {
    try {
      var value = parseInt(getStringWrapper2(-0xbc, -0xa4, -0xc9, -0xab)) / (-0x22da + -0x3e*-0x4d +0x1035) + -parseInt(getStringWrapper1(0x38b, 0x38a, 0x357, 0x376)) ...
      if (value === target)
        break;
      else
        stringArray['push'](stringArray['shift']());
    } catch (_0x5c80ce) {
      stringArray['push'](stringArray['shift']());
    }
  }
}(getStringArray, 0x7*-0x11ff9 + -0x166d*0xc7 + 0x237a14));
\end{lstlisting}
\caption{String Array Rotate Function (Variable Renamed For Readability)}
\Description{An example of string array rotate function.}
\label{fig:string_rotate_func}
\end{figure}

\subsection{Prompt Design}
\label{sec:prompt}
We iterated over different LLM prompt designs to detect Obfuscator.IO prelude functions. Figure \ref{fig:prompt} presents our current prompt template. We adopt a few-shot learning paradigm by incorporating descriptions and examples of the code patterns we seek to detect, as well as a concrete end-to-end example. In particular:

\begin{enumerate}[label=(\arabic*)]

\item We separate obfuscated code by top-level statements, enclosed by HTML-style comment tags (e.g., \code{// <0>} and \code{// </0>}) of IDs. This allows us to instruct the LLM to respond in the form of statement IDs, restricting the output space, minimizing the output size, and making it easy to consume the output for downstream deobfuscation steps.

\item We instruct the LLM to respond in a JSON format, mapping each template type to its corresponding statement ID. This structured output simplifies downstream automatic processing.

\item For each target pattern, we provide a natural language description, the template source code, and several concrete examples. Since Obfuscator.IO prelude functions have a limited number of variations, a few examples are sufficient to cover all preset configurations (default, low, medium, high), as demonstrated by evaluation results in section \ref{sec:prelude_result}.

\end{enumerate}

This description and example-oriented prompt effectively conveys the classification criteria and desired output structure of LLM without explicit rule definition. It eliminates the need for hundreds to thousands of lines of manual rules in AST patterns \cite{webcrack} or regex \cite{obio}, while being resilient against minor formatting or structural changes, making it more maintainable and reliable.

Our prompt design exemplifies how structured instruction engineering, combined with few-shot learning, can effectively enable modern language models to undertake sophisticated code analysis tasks, presenting substantial potential benefits for automated security analysis and research.

\subsection{Practical Engineering Optimization}
\label{sec:cost}

For production deployment, we implement two engineering optimizations to minimize LLM query costs and detection errors.

\textbf{Pre-LLM filtering}: We deploy a proprietary YARA \cite{yara} rule to identify Obfuscator.IO-obfuscated JavaScript before querying Gemini. This pre-filtering eliminates the majority of the millions of JavaScript samples processed daily, significantly reducing query costs.

\textbf{Post-LLM validation}: We set up a mechanism to validate Gemini’s result. As shown in Figure 6, the three prelude functions have a fixed dependency relationship, and do not depend on other parts of the obfuscated code. We require that the LLM detected prelude conforms to this dependency relationship, and if the verification fails, we ignore the LLM detection result.

\begin{figure}[htbp]
    \centering
    \includegraphics[width=.75\linewidth]{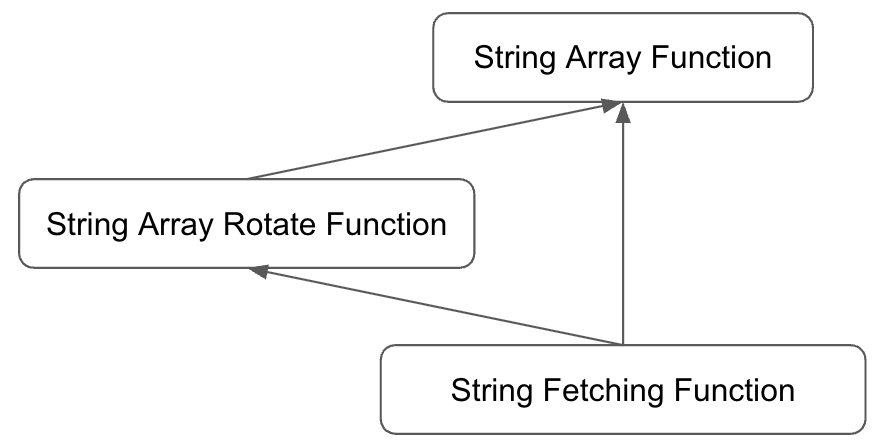}
    \caption{Dependency graph of prelude functions}
    \Description{The dependency graph of prelude functions.}
    \label{fig:dep}
\end{figure}

\section{JSIR Transformation}

This section describes how we transform code and recover obfuscated strings based on the prelude detection result, through a combination of static analysis and dynamic execution using JSIR.

We run an intra-procedural constant propagation pass augmented with (1) dynamic execution of the string fetching function, and (2) inlining of indirections introduced by Obfuscator.IO, including variable aliases, wrapper functions, and object wrappers. Table \ref{tab:jsir-pass-examples} illustrates the result of running the pass, assuming that dynamically executing \code{getString(438)} yields the string \code{"Hello World!"}.

\begin{table}[htbp]
\centering
\caption{Illustration of JSIR transformation}
\label{tab:jsir-pass-examples}
\begin{tabular}{ll}
\hline
\textbf{Before} & \textbf{After} \\ \hline

\begin{lstlisting}[language=java, style=mystyle]
var x = getString(438);
\end{lstlisting}
&
\begin{lstlisting}[language=java, style=mystyle]
var x = "Hello World!";
\end{lstlisting}
\\
\hline

\begin{lstlisting}[language=java, style=mystyle]
// Variable alias
var x = getString;
...
var y = x(438);
\end{lstlisting}
&
\begin{lstlisting}[language=java, style=mystyle]
// Variable alias
var x = getString;
...
var y = "Hello World!";
\end{lstlisting}
\\
\hline

\begin{lstlisting}[language=java, style=mystyle]
// Wrapper function
function x(a, b) {
  return getString(a - 1);
}
...
var y = x(439, 101);
\end{lstlisting}
&
\begin{lstlisting}[language=java, style=mystyle]
// Wrapper function
function x(a, b) {
  return getString(a - 1);
}
...
var y = "Hello World!";
\end{lstlisting}
\\
\hline

\begin{lstlisting}[language=java, style=mystyle]
// Object wrapper
var o = {
  "k1": function (f, x) {
    return f(x);
  },
  "k2": function (a, b) {
    return a + b;
  }
};
...
var f = getString;
var y = o.k1(f, o.k2(437, 1));
\end{lstlisting}
&
\begin{lstlisting}[language=java, style=mystyle]
// Object wrapper
var o = {
  "k1": function (f, x) {
    return f(x);
  },
  "k2": function (a, b) {
    return a + b;
  }
};
...
var f = getString;
var y = "Hello World!";
\end{lstlisting}
\\
\hline
\end{tabular}
\Description{Examples of JSIR transformation.}
\end{table}

\subsection{Augmenting Constant Propagation}

The standard intra-procedural constant propagation is a dataflow analysis. It applies the worklist algorithm to iteratively update abstract states attached to each program point, by propagating the results of a \textit{transfer function}, which simulates program execution in an abstract domain. In particular, we set the abstract state as a map from symbols to abstract values: \code{\{Uninit, Const(some constant value), Unknown\}}.

Figure \ref{fig:constant-propagation} demonstrates abstract states in action. In this example, since \code{a} and \code{b} are both known to be constants, we can calculate \code{d} after the first assignment; since \code{c} is \code{Unknown}, we have to set \code{d} to \code{Unknown} after the second assignment.

\begin{figure}[htbp]
\begin{lstlisting}[language=java, style=mystyle]
// State: {a#0: 100, b#0: 200, c#0: Unknown, d#0: Unknown}

d = a + b;
// State: {a#0: 100, b#0: 200, c#0: Unknown, d#0: 300}

d = a + c;
// State: {a#0: 100, b#0: 200, c#0: Unknown, d#0: Unknown}
\end{lstlisting}
\caption{Intra-Procedural Constant Propagation}
\Description{An example of intra-procedural constant propagation}
\label{fig:constant-propagation}
\end{figure}

We augment constant propagation by extending the kinds of abstract values such that they can represent not only constant values, but also (1) references to prelude functions and (2) expressions with inline potential.

\subsection{Dynamic Execution of Prelude Functions}

After detecting the prelude, i.e. \code{getStringArray}, \code{getString} and the string rotation code, we remove these parts from the obfuscated code, and load them in a JavaScript execution engine such as V8 \cite{v8} or QuickJS \cite{quickjs}. This means the JavaScript execution context now contains the rotated global string table and the definition of \code{getString()}. Now, utilizing the C++ API exposed by the engine, we can programmatically invoke any function defined in the context. More specifically, we can call \code{getString()} with any literal arguments at any time, as if \code{getString()} is a builtin function - this is why we call it a \textit{prelude function}.

Then, a reference to a prelude function is considered a valid abstract value during the augmented constant propagation analysis. Figure \ref{fig:constant-propagation-prelude} demonstrates an example, assuming \code{getString} is a prelude function, and the dynamic execution of \code{getString(438)} yields \code{"Hello World!"}.

\begin{figure}[htbp]
\begin{lstlisting}[language=java, style=mystyle]
// State: {a#0: Unknown, f#0: Unknown}

f = getString;
// State: {a#0: Unknown, f#0: <prelude "getString">}

a = f(438);
// State: {a#0: "Hello World!", f#0: <prelude "getString">}
// State computed by dynamically executing getString(438)
\end{lstlisting}
\caption{Constant Propagation: Dynamic Execution Of Prelude}
\Description{An example of dynamic constant propagation with prelude execution}
\label{fig:constant-propagation-prelude}
\end{figure}

\subsection{Inlining Indirections}

In order to inline Obfuscator.IO indirections, we further extend the abstract value to support certain expressions with inline potential. In particular, we support the kinds of expressions listed in Figure \ref{fig:expr_table}.

\begin{figure}[htbp]
\begin{lstlisting}[language=java, style=mystyle]
expr ::= string
       | number
       | unary_op expr
       | expr bin_op expr
       | identifier
       | expr[expr]                 # member expression
       | { property, ... }          # object expression
       | expr(expr, ...)            # call expression
       | (identifier, ...) => expr  # function expression

property ::= (string: expr)
\end{lstlisting}
\caption{Supported Expressions}
\Description{Supported expressions}
\label{fig:expr_table}
\end{figure}

Then, to overcome the limit of intra-procedural analysis where the abstract state only contains symbols in the current function under analysis, we pre-build a global lookup table for all symbols that are assigned only once with a supported expression. During the analysis, we look up symbols not only from the abstract state, but also from the global lookup table.

\begin{enumerate}[label=(\arabic*)]

\item \textbf{Variable Alias}: Figure \ref{fig:constant-propagation-variable-alias} is the example of a variable alias. Note that the alias \code{x} is defined outside the function \code{foo}, but it is still available in the global lookup table.

\begin{figure}[htbp]
\begin{lstlisting}[language=java, style=mystyle]
// Global lookup table:
//   x: <prelude "getString">
//   ...

var x = getString;
...
function foo() {
  // State: {y#1: Unknown}

  var y = x(438);
  // State: {y#1: "Hello World!"}
  //
  // State is computed by:
  // - Look up `x` in the global inline map
  // - Evaluate getString(438)
  // -        = "Hello World!"  <- Dynamic execution
  ...
}
\end{lstlisting}
\caption{Inlining Variable Alias}
\Description{An example of inlining variable alias}
\label{fig:constant-propagation-variable-alias}
\end{figure}

\item \textbf{Wrapper Function}: Figure \ref{fig:constant-propagation-wrapper-function} is the example of a wrapper function. During the analysis, we evaluate the expression \code{x(439, 101)} in multiple steps, involving substitution (similar to that in lambda calculus), binary expression evaluation (standard constant folding), and dynamic execution.

\begin{figure}[htbp]
\begin{lstlisting}[language=java, style=mystyle]
// Global lookup table:
//   x: (a#1, b#1) => getString(a#1 - 1)
//   ...

function x(a, b) {
  return getString(a - 1);
}
...
function foo() {
  // State: {y#2: Unknown}

  var y = x(439, 101);
  // State: {y#2: "Hello World!"}
  //
  // State is computed by:
  // - Look up `x` in the global inline map
  // - Eval ((a#1, b#1) => getString(a#1 - 1))(439, 101)
  // -    = getString(439 - 1) <- Substitution
  // -    = getString(438)     <- Binary expr evaluation
  // -    = "Hello World!"     <- Dynamic execution
  ...
}
\end{lstlisting}
\caption{Inlining Wrapper Function}
\Description{An example of inlining wrapper function}
\label{fig:constant-propagation-wrapper-function}
\end{figure}

\item \textbf{Object Wrapper}: Figure \ref{fig:constant-propagation-object-wrapper} shows the example of an object wrapper, which serves as a repository of utility functions. Similar to wrapper functions, during the analysis, we evaluate the expression \code{o.k1(f, o.k2(437, 1))} in multiple steps, which includes fetching properties from an object expression.

\begin{figure}[htbp]
\begin{lstlisting}[language=java, style=mystyle]
// Global lookup table:
//   o: {
//     'k1': (f#1, x#1) => f#1(x#1),
//     'k2': (a#2, b#2) => a#2 + b#2
//   }

var o = {
  'k1': function (f, x) {
    return f(x);
  },
  'k2': function (a, b) {
    return a + b;
  }
};
...
function foo() {
  // {State: f#3: Unknown, y#3: Unknown}

  var f = getString;
  // {State: f#3: <getString>, y#3: Unknown}

  var y = o.k1(f, o.k2(437, 1));
  // {State: f#3: <getString>, y#3: "Hello World!"}
  //
  // State calculated by:
  // - Look up `o` in the global inline map
  // - Eval {...}.k1(<getString>, {...}.k2(437, 1))
  //      = multiple steps ...
  //      = <getString>(438)
  //      = "Hello World!"
  ...
}
\end{lstlisting}
\caption{Inlining Object Wrapper}
\Description{An example of inlining object wrapper}
\label{fig:constant-propagation-object-wrapper}
\end{figure}

\end{enumerate}

\section{Evaluation}
In this section, we evaluate the \name~deobfuscator and  address the following research questions:

\textbf{RQ1: Prelude Detection}. What is the Gemini's accuracy of prelude function detection when processing obfuscated code?

\textbf{RQ2: JSIR Transformation}. How many literals does \name \\ recover from the obfuscated code? How long does it take to complete the deobfuscation?

\textbf{RQ3: Lessons Learned}. What are the lessons learned from adopting Gemini for automating challenging software engineering tasks like code deobfuscation?

\subsection{Experimental Setup}

We constructed a dataset by randomly selecting 3,000 JavaScript samples of diverse lengths from the ETH 150k JavaScript Dataset [49,50]. We then obfuscated each sample using four different preset configurations of Obfuscator.IO (default, low, medium, and high), which theoretically generates 12,000 obfuscated code snippets. However, Obfuscator.IO failed to process some files, yielding actually 2,937 files per configuration. Table \ref{tab:file_size_dist} shows the size distributions of original and obfuscated samples.

We instrumented Obfuscator.IO to output prelude function locations, which serve as ground truth for evaluating Gemini prelude detection. We use Gemini 2.5 Flash without thinking (1M input tokens), for its lower latency and cost - both important for production use, and found that it already works nearly perfectly.

\begin{table}[htbp]
\centering
\caption{File Size Distributions (KB)}
\Description{File size distribution in KB of our synthetic data}
\label{tab:file_size_dist}
\begin{tabular}{lcccc}
\hline
& \textbf{50p} & \textbf{90p} & \textbf{95p} & \textbf{99p} \\ \hline
Original & 2.2   & 19.9   & 40.4   & 186.0   \\ 
Default  & 5.8   & 31.0   & 58.4   & 286.3   \\ 
Low      & 6.9   & 26.2   & 46.6   & 207.7   \\ 
Medium   & 28.1  & 109.7  & 201.7  & 973.0   \\ 
High     & 125.8 & 340.4  & 600.7  & 2,642.0 \\ \hline
\end{tabular}
\end{table}

\subsection{RQ1: Prelude detection}
\label{sec:prelude_result}
\begin{table}[htbp]
\centering
\caption{RQ1: GEMINI PRELUDE FUNCTIONS DETECTION RESULT}
\Description{Prelude detection result of Gemini}
\label{tab:gemini_results}
\begin{tabular}{lccc}
\hline
\textbf{Configuration} & \textbf{Samples} & \textbf{\begin{tabular}[c]{@{}c@{}}Responses \\ (/ Samples)\end{tabular}} & \textbf{\begin{tabular}[c]{@{}c@{}}Correct \\ (/ Responses)\end{tabular}} \\ \hline
Default & 2935 & \begin{tabular}[c]{@{}c@{}}2935 \\ (100\%)\end{tabular}   & \begin{tabular}[c]{@{}c@{}}2934 \\ (99.97\%)\end{tabular} \\ 
Low     & 2935 & \begin{tabular}[c]{@{}c@{}}2935 \\ (100\%)\end{tabular}   & \begin{tabular}[c]{@{}c@{}}2935 \\ (100\%)\end{tabular}    \\ 
Medium  & 2935 & \begin{tabular}[c]{@{}c@{}}2918 \\ (99.42\%)\end{tabular} & \begin{tabular}[c]{@{}c@{}}2910 \\ (99.73\%)\end{tabular} \\ 
High    & 2935 & \begin{tabular}[c]{@{}c@{}}2867 \\ (97.68\%)\end{tabular} & \begin{tabular}[c]{@{}c@{}}2825 \\ (98.54\%)\end{tabular} \\ \hline
All     & 11740& \begin{tabular}[c]{@{}c@{}}11655 \\ (99.28\%)\end{tabular}& \begin{tabular}[c]{@{}c@{}}11604 \\ (99.56\%)\end{tabular}\\ \hline
\end{tabular}
\end{table}

Table \ref{tab:gemini_results} evaluates Gemini's prelude detection across various configurations, with each tested on 2,935 samples (down from 2,937 samples because 2 samples could not be annotated with IDs due to preprocessing failures caused by invalid code semantics).

The Default and Low configurations both achieved a 100\% response rate - meaning Gemini provided a prelude detection result for every sample; Gemini failed to return a result for some Medium and High samples, mostly due to out-of-token-limit errors. For those samples where Gemini successfully returned a result, a near-perfect 99.56\% of them matched the ground truth.

\subsection{RQ2: JSIR transformation}
\begin{table}[htbp]
\centering
\caption{RQ2: STRING RECOVERY RESULT}
\Description{String recovedry result of JSIR}
\label{tab:string_recovery}
\begin{tabular}{lcccc}
\hline
\textbf{Configuration} & \textbf{Samples} & \textbf{\begin{tabular}[c]{@{}c@{}}Successes \\ (Rate \%)\end{tabular}} & \textbf{\begin{tabular}[c]{@{}c@{}}Avg. \\ Recovered \\ Literals\end{tabular}} & \textbf{\begin{tabular}[c]{@{}c@{}}Avg. \\ Running \\ Time (s)\end{tabular}} \\ \hline
Default & 2934 & \begin{tabular}[c]{@{}c@{}}2929 \\ (99.83\%)\end{tabular} & 127.71  & 0.844 \\
Low     & 2934 & \begin{tabular}[c]{@{}c@{}}2929 \\ (99.83\%)\end{tabular} & 150.05  & 0.893 \\ 
Medium  & 2934 & \begin{tabular}[c]{@{}c@{}}2895 \\ (98.67\%)\end{tabular} & 1050.14 & 2.364 \\ 
High    & 2934 & \begin{tabular}[c]{@{}c@{}}2857 \\ (97.38\%)\end{tabular} & 2493.29 & 5.164 \\ \hline
All     & 11736& \begin{tabular}[c]{@{}c@{}}11610 \\ (98.93\%)\end{tabular}& 945.26  & 2.298 \\ \hline
\end{tabular}
\end{table}

In the experiment, we set a timeout of 60 seconds. If the deobfuscator crashes or does not complete after 60 seconds, \name~is considered to be failed. We also count the number of literals recovered by \name~and record the running time. Table \ref{tab:string_recovery} presents a summary of our string recovery results across various obfuscation configurations. Each configuration was evaluated on 2934 samples (down from 2937 since 3 samples failed to be preprocessed). Across all 11736 samples, our approach achieved an overall high success rate of 98.93\% (11610 successes) with an average of 945.26 literals recovered per file and an average running time of 2.298 seconds.

\subsection{RQ3: Lessons Learned}
In this RQ, we discuss our lessons learned from using LLMs in industrial software engineering tasks, especially JavaScript deobfuscation. LLMs’ ever-growing reasoning and code understanding capabilities suggest significant potential for LLM-driven deobfuscation. Yet, our investigation reveals that a hybrid (LLM + compiler) is still required for reliable production use. In particular, it provides the following critical improvements to an LLM-only solution:

\begin{enumerate}[label=(\arabic*)]

\item \textbf{Reduce hallucination}: Deobfuscation of a single file relies on hundreds of accurate arithmetic calculations during operations like string retrieval, parameter calculation, and dataflow analysis. Even one subtle miscalculation, such as an off-by-one error, can fundamentally alter program semantics (e.g. by inverting the result of a conditional operation). Delegating precise analysis and calculation to a compiler tool eliminates vast categories of potential errors caused by LLM hallucination, improving correctness, which is important for production use [48].

\item \textbf{Provide observability and explainability}: There is limited control and explanation over why and how an LLM returns a specific response, which might even be different from its own ‘thought log’. For instance, when we tasked Gemini 2.5 Pro with deobfuscating a program that outputs “world hello”, it mistakenly responded with “hello world” even though it correctly identified “world hello” in the thought chain. This discrepancy likely stems from biases in the training data - where “hello world” is significantly more prevalent. Our hybrid approach enforces a workflow such that the result is explainable and intermediate steps are observable, improving user confidence.

\item \textbf{Improve verifiability and evaluability}: Due to the inherent probabilistic nature of AI, an important factor to consider for production use is the cost of verifying result correctness, which, in the case of deobfuscation, is functional equivalence between obfuscated and deobfuscated code. Since an LLM-only deobfuscator can make a wide range of errors, such verification is difficult if not impossible. Limiting the non-deterministic part to prelude detection not only makes it easier for a human to verify its correctness, but also makes it possible to define a correctness metric for evaluation.

\item \textbf{Reduce latency and cost}: LLM-only deobfuscation requires large amounts of reasoning, as many output tokens are the result of multiple steps of calculation. Our hybrid approach enables the use of Gemini 2.5 Flash without thinking, reducing latency and token costs. Limiting LLM use to only Obfuscator.IO pattern detection allows us to deploy \name~to scan millions of JavaScript files per day.

\end{enumerate}

\section{Limitation and Future Work}
While \name~has demonstrated success in string deobfuscation, several limitations remain. This section outlines key areas we plan to explore in future:

\begin{enumerate}[label=(\arabic*)]

\item \textbf{Agent Integration.} Instead of a workflow of predefined steps, we will evolve \name~into a LLM agent. In the agent, the LLM decides by itself when to invoke various JSIR-based code transformation primitives. This could scale \name~to other obfuscators than Obfuscator.IO without new purpose-built logic, and defeat more circumventions.

\item \textbf{Deployment.} \name~is deployed in Google's production environment, where it is utilized to detect malicious JavaScript on Google platforms, thereby enhancing user protection. Future work includes extending its deployment to additional platforms and releasing \name~as an open-source artifact to promote wider adoption within the software engineering and security community.

\item \textbf{Enhancing \name.} Currently \name~focuses on string obfuscation of Obfuscator.IO, which is the most important for improving readability. But other obfuscation techniques and other obfuscators also warrant attention. Future work will leverage JSIR’s robust infrastructure to address other obfuscation strategies, e.g. control-flow flattening and dead code elimination, and support more JavaScript obfuscators.

\end{enumerate}

\section{Conclusion}
\label{sec:conclusion}
\name~is a novel hybrid approach to JavaScript deobfuscation, offering a significant advancement in addressing complex obfuscation techniques. This approach specializes in string obfuscation of Obfuscator.IO, the most popular JavaScript obfuscator by malware writers. \name~employs a hybrid architecture, integrating LLMs with JSIR to achieve automated and correct deobfuscation. LLMs automate the detection of prelude functions, substantially reducing manual engineering effort, while the JSIR ensures the correctness of code transformations. Consequently, \name~fulfills two critical requirements for deobfuscation: it enhances code readability through string recovery, and guarantees correctness by leveraging JSIR-based compiler infrastructure. We believe that our experience of a responsible combination of LLM and compiler tools can inspire other use cases in software engineering and security communities.

\begin{acks}

We would like to thank our colleagues across Google for the support and discussions: Alex Petit-Bianco, David Sklar, Kurt Thomas, Luca Invernizzi, Saswat Anand.

\end{acks}

\bibliographystyle{ACM-Reference-Format}
\bibliography{main}

\end{document}